\documentclass[pra,aps,showpacs,groupedaddress,superscriptaddress,twocolumn,toc=flat]{revtex4-1}

\usepackage[colorlinks=true,linkcolor=blue,urlcolor=blue,citecolor=blue]{hyperref}

\usepackage[utf8x]{inputenc}
\usepackage{color}
\usepackage{bbm} 

\usepackage{amsfonts,amsmath,amssymb,stmaryrd}

\usepackage{graphicx}
\usepackage{subfigure}  
\usepackage{bbm} 
\usepackage{hyperref}
\usepackage{epsfig}
\usepackage{mathrsfs}
\usepackage{verbatim}
\usepackage{centernot}
\usepackage{ulem}
\usepackage{nicefrac}
\usepackage{xspace}
\usepackage{siunitx}
\usepackage{xfrac}

\renewcommand{\l}{\left(}
\renewcommand{\r}{\right)}

\newcommand{\ket}[1]{\ensuremath{\lvert #1 \rangle}\xspace}%
\newcommand{\avg}[1]{\ensuremath{\langle #1 \rangle}\xspace}%
\renewcommand{\ij}{{\langle \vec{i}, \vec{j} \rangle}}

\renewcommand{\H}{\hat{\mathcal{H}}}

\renewcommand{\a}{\hat{a}}

\newcommand{\ad}{\hat{a}^\dagger}

\newcommand{\hc}{\text{h.c.}}

\usepackage{array}

\usepackage{cancel,ifthen}
\newcommand{\cmnt}[2][NoInPuT]{\ifthenelse{\equal{#1}{NoInPuT}}{}{{\color{red}\sout{#1}}} {\color{blue} #2}}

\usepackage{bm}	
\renewcommand{\vec}[1]{\bm{#1}}

\begin{document}
\normalem	

\title{Observation of constrained dynamics of bosonic charge carriers in\\ staggered magnetic fields}
\title{Microscopy of bosonic charge carriers in staggered magnetic fields}

\author{Annabelle Bohrdt}
\thanks{These authors contributed equally.}
\affiliation{University of Regensburg, Universitätsstr. 31, Regensburg D-93053, Germany}
   \affiliation{Munich Center for Quantum Science and Technology (MCQST), 80799 Munich, Germany}

\author{David Wei}
\thanks{These authors contributed equally.}
   \affiliation{Munich Center for Quantum Science and Technology (MCQST), 80799 Munich, Germany}
    \affiliation{Max-Planck-Institut f\"{u}r Quantenoptik, 85748 Garching, Germany}

\author{Daniel Adler}   
    \affiliation{Munich Center for Quantum Science and Technology (MCQST), 80799 Munich, Germany}
    \affiliation{Max-Planck-Institut f\"{u}r Quantenoptik, 85748 Garching, Germany}

\author{Kritsana Srakaew}
    \affiliation{Munich Center for Quantum Science and Technology (MCQST), 80799 Munich, Germany}
    \affiliation{Max-Planck-Institut f\"{u}r Quantenoptik, 85748 Garching, Germany}

\author{Suchita Agrawal}
    \affiliation{Munich Center for Quantum Science and Technology (MCQST), 80799 Munich, Germany}
    \affiliation{Max-Planck-Institut f\"{u}r Quantenoptik, 85748 Garching, Germany}

\author{Pascal~Weckesser}
    \affiliation{Munich Center for Quantum Science and Technology (MCQST), 80799 Munich, Germany}
    \affiliation{Max-Planck-Institut f\"{u}r Quantenoptik, 85748 Garching, Germany}

    \author{Immanuel Bloch}
    \affiliation{Munich Center for Quantum Science and Technology (MCQST), 80799 Munich, Germany}
    \affiliation{Max-Planck-Institut f\"{u}r Quantenoptik, 85748 Garching, Germany}
    \affiliation{Fakult\"{a}t f\"{u}r Physik, Ludwig-Maximilians-Universit\"{a}t, 80799 Munich, Germany}

\author{Fabian Grusdt}
    \affiliation{Munich Center for Quantum Science and Technology (MCQST), 80799 Munich, Germany}
    \affiliation{Department of Physics and Arnold Sommerfeld Center for Theoretical Physics (ASC), Ludwig-Maximilians-Universit\"at M\"unchen, Theresienstr. 37, M\"unchen D-80333, Germany}

\author{Johannes Zeiher}  
    \affiliation{Munich Center for Quantum Science and Technology (MCQST), 80799 Munich, Germany}
    \affiliation{Max-Planck-Institut f\"{u}r Quantenoptik, 85748 Garching, Germany}

\date{\today}

\begin{abstract}
The interplay of spin and charge degrees of freedom is believed to underlie various unresolved phenomena in strongly correlated systems. Quantum simulators based on neutral atoms provide an excellent testbed for investigating such phenomena and resolving their microscopic origins. Up to now, the majority of experimental and theoretical studies has focused on systems with fermionic exchange statistics. Here we expand the existing cold atom toolbox through the use of negative temperature states, enabling us to realize an antiferromagnetic, \textit{bosonic} $t-J$ model in two spatial dimensions, subject to a strong staggered magnetic field in a quantum gas microscope. Through comparison of the spreading dynamics of a single hole in a N\'eel versus a spin-polarized initial state, we establish the relevance of memory effects resulting from the buildup of strong spin-charge correlations in the dynamics of charge carriers in antiferromagnets. We further numerically predict rich dynamics of pairs of doped holes, which we demonstrate to be bound by a similar memory effect, while their center-of-mass can expand freely. Our work paves the way for the systematic exploration of the effect of antiferromagnetic spin ordering on the properties of individual charge carriers as well as finite doping phases: Our study demonstrates that the staggered field can be used to single out the effect of antiferromagnetism and holds the prospect to prepare low-temperature states in the near future.
\end{abstract}

\maketitle

\section{Introduction}\label{sec:intro}
The microscopic properties of individual charge carriers as well as pairs of charge carriers in doped quantum antiferromagnets (AFMs) constitute a key stepping stone in the quest to understand unconventional superconductivity. Properties of interest encompass the local spin surrounding of an individual hole~\cite{Grusdt2019,Koepsell2019,Nielsen2021} as well as pairs of holes~\cite{White1997}, the distance between two dopants~\cite{Blomquist2021}, and their propagation speed~\cite{Bohrdt2020_dynamics,Bohrdt2020_ARPES,Grusdt2022,Bohrdt2023}. 
Solid state experiments typically access frequency and momentum-resolved information, thus, e.g. providing information on the dispersion relation of a single dopant through angle-resolved photoemission spectroscopy (ARPES)~\cite{Wells1995,Brunner2000,Mishchenko2001,Ronning2005,Graf2007,Manousakis2007,Bohrdt2020_ARPES}. Probing properties of pairs directly in materials is challenging, and insights can typically either be obtained indirectly, or at a low signal-to-noise ratio, e.g. through coincidence ARPES~\cite{Berakdar1998,Su2020,Mahmood2022} or noise spectroscopy~\cite{Bastiaans2021}.  

Microscopic studies of dopants in real time and space provide a complementary picture to spectroscopic probes. However, the numerical simulation of the time evolution of a strongly correlated, two-dimensional quantum system is by itself challenging. While solid-state experiments often cannot directly access real-time and spatial properties simultaneously, quantum simulators using cold atoms in optical lattices are ideally suited to study the dynamics of itinerant particles in real time and space~\cite{Gross2017,Bohrdt2021_review}. Experiments using cold fermionic atoms have, for example, demonstrated dynamical spin-charge separation in one dimension~\cite{Kollath2005,Vijayan2020}, as well as the formation and dynamical spreading of a magnetic polaron after releasing a single hole in the two-dimensional (2D) Fermi-Hubbard model~\cite{Ji2021}. 

\begin{figure*}
\centering
\includegraphics{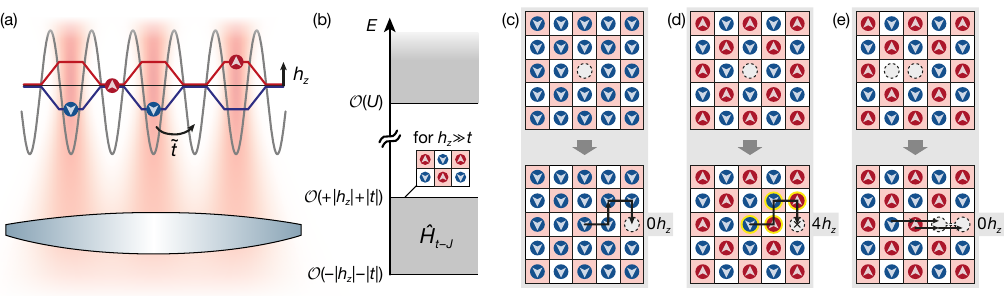} 
\caption{
\textbf{Experimental realization of an AFM, bosonic $t-J-V$ model in a strong staggered magnetic field} $\pm h_z$. (a) We trap ultracold bosonic $^{87}\mathrm{Rb}$ atoms in two hyperfine states in an optical lattice and realize the staggered field by projecting spin-differential ``anti-magic'' local light shifts through a high-numerical-aperture microscope objective. (b) By realizing negative absolute temperature states at the upper end of the spectrum, but in the subspace without double occupancies, we are able to explore the physics of the bosonic AFM $t-J-V$ model. To this end we initialize a single, localized hole in the center of the system and analyze its subsequent spreading dynamics: (c) In a spin-polarized background, the hole expands without leaving a memory of its trajectory and with the staggered field representing an effective staggered potential. (d) In the AFM N\'eel background, the hole motion imprints a strong memory of its trajectory and the staggered magnetic field creates a string tension $\propto h_z$ hindering free hole motion. Theoretically we also analyze initial states (e) with two holes localized next to each other in a N\'eel background. The pair is found to be bound together by the staggered field while its center-of-mass expands freely, resembling the motion of a single hole in the spin-polarized background (c). 
}
\label{fig:negativeT}
\end{figure*}

Here we extend these studies including the use of bosonic atoms to simulate an antiferromagnetic spin system~\cite{Trotzky2008,Jepsen2020,Sun2021} and the addition of a staggered magnetic field~\cite{Yang2020}, which provides a new tuning knob to control the interplay of spin and charge degrees of freedom, see Fig.~\ref{fig:negativeT}(a). We demonstrate experimentally that for the dynamics of a single hole, the initial state of the spin background plays a crucial role: In a N\'eel-state spin background, the free motion of the hole is found to be inhibited. We explain this by a memory of the hole's trajectory, which is imprinted into the spin background in the form of a string of displaced spins~\cite{Trugman1988,Simons1990}, see Fig.~\ref{fig:negativeT}(d). In contrast, there is no such memory effect for a hole moving in a spin-polarized initial state, see Fig.~\ref{fig:negativeT}(c), and the hole can spread through the system freely despite the staggered potential landscape, as we confirm experimentally. Our work thus represents, to our knowledge, the first microscopic experimental study of few-hole dynamics in doped bosonic Hubbard systems in 2D.

Going beyond the presently accessible regime in our experiment, we theoretically analyze the dynamics of two holes with bosonic exchange statistics, initialized on neighboring lattice sites in the presence of a strong staggered field, see Fig.~\ref{fig:negativeT}(e). Similar to the motion of a single hole in a spin-polarized background, there is an intermediate state, when only one hole has moved, with an energy proportional to the strength of the staggered magnetic field. In this case, the second hole can retrace the string created by the first hole, enabling the pair to spread through the lattice freely without a memory of its trajectory.


The memory left behind due to the motion of the hole can be probed through string patterns. While a similar string mechanism has been argued to be in place in SU(2) invariant $t-J$ and Hubbard models~\cite{Liu1991,Mishchenko2001,Manousakis2007,Bohrdt2020_ARPES,Bohrdt2021PRL}, the analysis of string patterns is challenging: In this case the signal-to-noise ratio is comparably low~\cite{Chiu2019Science,Bohrdt2020_dynamics} due to significant quantum fluctuations for measurements in the $S^z$ basis. In contrast, the staggered magnetic field can be used experimentally to control the degree of quantum fluctuations. For sufficiently strong fields, these fluctuations are suppressed, and the string of displaced spins left behind by a hole moving in a N\'eel spin background can be directly observed in each snapshot, each corresponding to a quantum-projective measurement in the Fock basis.

Additionally, the strength of the staggered field provides a tuning knob to control properties of a single and two holes, such as for example their binding energy and the pair size. Beyond individual dopants, the question arises how a staggered magnetic field changes the entire phase diagram of $t-J$ or Fermi-Hubbard type models: Although antiferromagnetism is believed to play an essential role for the ensuing low-energy physics, its quick disappearance with doping remains poorly understood. In the setting explored in this article, AFM ordering of the spin sector can be directly controlled, enabling systematic studies of the phase diagram in the future. Further motivation for such studies comes from the ability to use the staggered field to adiabatically prepare low-temperature N\'eel states without crossing a phase transition. 

The remainder of this paper is organized as follows: We start by introducing the model under consideration in Sec.~\ref{sec:model}. We discuss its experimental realization in a quantum gas microscope for in Sec.~\ref{sec:exp_real} and show results for the dynamics of a single hole in Sec.~\ref{sec:exp_results}. In Sec.~\ref{sec:pair}, we theoretically analyze the dynamics of a pair of charge carriers in this setting, before we close by providing an outlook in Sec~\ref{sec:summary}.


\section{AFM bosonic $t-J-V$ model}\label{sec:model}
The starting point for our analysis is the two-dimensional $t-J-V$ model with an external staggered magnetic field of strength $h_z$:
\begin{equation}
    \begin{split}
\H_{t-J-V} = &- t ~ \hat{\mathcal{P}} \sum_{\ij} \sum_\alpha \l  \ad_{\vec{i},\alpha} \a_{\vec{j},\alpha} + \hc \r \hat{\mathcal{P}} + \\
&+ J \sum_{\ij} \hat{\vec{S}}_{\vec{i}} \cdot \hat{\vec{S}}_{\vec{j}} +V \sum_{\ij} \hat{n}_{\vec{i}} \hat{n}_{\vec{j}} + \\
&+  \frac{h_z}{2} \sum_{\vec{i}} (1+(-1)^{i_x+i_y}) \hat{S}_{\vec{i}}^z,
\label{eqtJModel}
\end{split}
\end{equation}
where $\hat{\mathcal{P}}$ projects to the subspace with maximum single occupancy per site; $\a^{(\dagger)}_{\vec{j},\alpha}$ are the annihilation (creation) operators on site $\vec{j}$ and with spin index $\alpha=\uparrow,\downarrow$, $\hat{\vec{S}}_{\vec{j}}$ and $\hat{n}_{\vec{j}}$ denote the on-site spin and density operators, respectively. We assume {\it{antiferromagnetic}} spin-exchange interactions $\propto J >0$ and specify the strength $V$ of nearest neighbor density-density interactions below.
The increased energy cost of displacing spins from their favoured sublattice, due to the staggered magnetic field in this model, breaks the ${\rm SU}(2)$ symmetry down to ${\rm U}(1)$. This renders its low-energy physics more similar to the $t-J_z$ model~\cite{Chernyshev1999} and facilitates a direct comparison to effective analytical models of its dopants in terms of strings~\cite{Bulaevskii1968,Trugman1988,Grusdt2018tJz}. 

In condensed matter theory, the $t-J$ models are conventionally studied as the strong interaction limit of the Fermi-Hubbard model, and thus the creation and annihilation operators $\a^{(\dagger)}_{\vec{j},\alpha}$ normally obey fermionic statistics. Here, instead, we consider a {\it{bosonic}} $t-J-V$ model, i.e.
\begin{equation}
    [\a_{\vec{j},\alpha},\a^{\dagger}_{\vec{j},\alpha'}] = \delta_{\vec{i},\vec{j}} \delta_{\alpha,\alpha'},
\end{equation}
notably, still with AFM spin-exchange $J>0$. We have realized this model using ultracold bosons in an optical lattice as discussed in the next section. Alternative implementations of the AFM, bosonic $t-J-V$ model have also been proposed, using dipolar atoms~\cite{Gorshkov2011tJ,Homeier2024} or Rydberg excitations in tweezer arrays~\cite{Homeier2024}, and related models have been realized experimentally~\cite{Sun2021,Jepsen2020,Jepsen2021} and discussed theoretically~\cite{Boninsegni2008,Aoki2009,Nakano2011,Nakano2012,Dicke2023}.

Importantly, for up to a single dopant, the statistics do not matter, and the bosonic and fermionic versions of the $t-J$ model are identical. However in the case of more than one dopant, or finite doping, the statistics generally makes a difference. The comparison between fermionic and bosonic models potentially enables crucial insights into the relevance of fermionic statistics in the interplay between charge and spin degrees of freedom for the realization of e.g. striped or superconducting phases~\cite{Smakov2004}. Thus far, little is known about the low energy phase diagram of the antiferromagnetic bosonic $t-J$ model, as most previous studies have considered lower dimensions, high temperature expansions, or only partial AFM couplings ($J_z>0, J_\perp \leq 0$)~\cite{Smakov2004,Boninsegni2008,Aoki2009,Nakano2011,Nakano2012,Dicke2023}. Recent density matrix renormalization group calculations of the ground state of the AFM, bosonic $t-J$ model with two dopants show a tendency towards the formation of a minimal stripe~\cite{Homeier2024}, with a numerical exploration of the finite doping regime forthcoming~\cite{Harris2024}.

In this paper, we consider a strong staggered magnetic field, which potentially attenuates the differences between fermionic and bosonic charge carriers.

\section{Experimental realization}\label{sec:exp_real}
In quantum simulation experiments, Hubbard or $t-J$ type models featuring AFM spin-exchange couplings are most naturally realized using fermionic atoms in optical lattices~\cite{Tarruell2018,Bohrdt2021_review}. 
The possibility to use neutral bosonic atoms instead for the study of quantum magnetism has been pointed out early on~\cite{Kuklov2003,Duan2003,GarciaRipoll2003,Altman2003}, and led to the first realization of coherent magnetic exchange couplings in quantum simulators~\cite{Trotzky2008}. The use of bosonic $^{87}\mathrm{Rb}$ atoms, in particular, has the advantage that staggered magnetic fields can be readily implemented with locally controlled differential light shifts, see Fig.~\ref{fig:negativeT}. The same technique also allows for the preparation of product N\'eel states in the 2D lattice.


Our experimental study employs a quantum gas microscope of $^{87}\mathrm{Rb}$ atoms, which supports single-site addressing techniques to manipulate the hyperfine states of individual atoms~\cite{Weitenberg2011,Fukuhara2013}.
Our method involves the projection of light fields which induce a differential light shift between the pseudospin states $\ket{\downarrow} = \ket{F=1, m_F=-1}$ and $\ket{\uparrow} = \ket{F=2, m_F=-2}$, such that a subsequent global microwave transfer pulse is only resonant to selected atoms~\cite{Weitenberg2011}.
By choosing addressing light at a wavelength of $\SI{788.7}{nm}$ with $\sigma^+$ polarization, we furthermore realize an ``antimagic'' condition, where the light shift of one spin state is opposite to the shift experienced by the other spin state.
Working in a bow-tie optical lattice with a spacing of $\SI{752}{nm}$ allows us to generate site-programmable local fields; in particular, this allows us to realize strong staggered magnetic fields with spatial inhomogeneities below $\SI{6}{\percent}$, see Appendix~\ref{sec:exp-details}.

We work in a sufficiently deep optical lattice to realize strong Hubbard-type interactions $U_{\alpha \alpha'}$ between the atoms, leading to a Mott insulator if a state with one boson per site is prepared. In our experiments, we implement the following spin-$1/2$ Bose-Hubbard model in a staggered field, see Fig.~\ref{fig:negativeT}(a),
\begin{multline}
    \H = - t ~ \sum_{\ij} \sum_\alpha \l  \ad_{\vec{i},\alpha} \a_{\vec{j},\alpha} + \hc \r  
+ U_{\uparrow \downarrow} \sum_{\vec{j}}  \hat{n}_{\vec{j},\uparrow} \hat{n}_{\vec{j},\downarrow} + \\
+ \sum_{\alpha}\frac{U_{\alpha\alpha}}{2} \sum_{\vec{j}} \hat{n}_{\vec{j},\alpha} (\hat{n}_{\vec{j},\alpha}-1) - \\
-  \frac{g_z}{4} \sum_{\vec{j}} (1+(-1)^{j_x+j_y}) (\hat{n}_{\vec{j},\uparrow} - \hat{n}_{\vec{j},\downarrow}).
\label{eqBHModel}
\end{multline} 
Typical experimental parameters are $t = h \times \SI{11.8(5)}{Hz}$ and $U_{\uparrow \downarrow} \approx U_{\uparrow \uparrow} \approx U_{\downarrow \downarrow} \equiv U = h \times \SI{250(20)}{Hz}$, with a freely tunable staggered field $g_z$.

We work in a regime with up to one boson per lattice site, as justified by the strong Hubbard interactions. The lowest-energy states of the Bose-Hubbard model~\eqref{eqBHModel} can thus be described by the bosonic $t-J-V$ model~\eqref{eqtJModel} with Hamiltonian $\H_{-}$, however, with ferromagnetic spin-exchange couplings $J_-= - 4 t^2/ U <0$~\cite{Duan2003}; other parameters are $h_{z-} = -g_z$ and $V_-=\sfrac{3}{4} J_-<0$, i.e. neighboring holes attract each other. Stable ground states with AFM spin correlations, despite the ferromagnetic couplings $J_-$, can be obtained for sufficiently strong fields: For $|h_{z-}| \gtrsim 4 |J_-|$ the N\'eel state with $S^z_{\vec{j}} = (-1)^{j_x+j_y} {\rm sgn}(g_z)$, i.e. aligned with the staggered field, corresponds to the overall ground state of the system. 

Although we considered the bosonic, ferromagnetic $t-J-V$ model $\H_{-}$ as a point of comparison in our experiments, see Appendix~\ref{secAppendix}, our primary goal is to study the bosonic $t-J-V$ model~\eqref{eqtJModel} with AFM couplings, $J>0$. In order to access this regime, we work with the highest-energy states of the ferromagnetic $t-J-V$ model $\H_{-}$ introduced above. Experimentally we realize a negative absolute-temperature state~\cite{GarciaRipoll2003,Sorensen2010,Braun2013} $\hat{\rho} = e^{-\beta_- \H_{-}} /Z$ with $1/\beta_- = k_B T_- < 0$, in the Hilbert space with no more than one boson per lattice site. This last condition is guaranteed by the strong repulsive Hubbard interactions, representing the largest energy scale in our system. We initialize the system using single-site addressing to prepare a product N\'eel state with 
\begin{equation}
    S^z_{\vec{j}} = -(-1)^{j_x+j_y} {\rm sgn}(g_z),
    \label{eqDefNeel}
\end{equation}
i.e. anti-aligned with the staggered field.

The high-energy physics of the negative temperature state we prepare can be equivalently described by a positive temperature state of the bosonic AFM $t-J-V$ model, $\hat{\rho} = e^{-\beta \H_{t-J-V}}/Z$. Here $1/\beta = k_B |T_-| > 0$ and $\H_{t-J-V}=-\H_{-}$, i.e., we realized the desired bosonic Hamiltonian Eq.~\eqref{eqtJModel} with 
\begin{equation}
    J=4 t^2/U > 0, \qquad h_z=g_z,
\end{equation}
and a nearest-neighbor repulsion $V=\sfrac{3}{4} J > 0$ between holes. In this process, the sign of the hopping also switches, $t \to - t$; However, on the bi-partite lattice we consider, ${\rm sgn}(t)$ can be changed by a gauge transformation ($\a_{\vec{j}} \to -\a_{\vec{j}}$ on only one sub-lattice), leaving the physics invariant.

In the following, we analyze the dynamics of a single hole, or a pair of dopants, by initializing the system with up to one boson per site and working in the highest energy state of $\H_-$, i.e., the lowest-energy state of the effective Hamiltonian $\H_{t-J-V}$. This approach is very versatile, and allows e.g. to use adiabatic ramps following the upper edge of the $t-J-V$ spectrum in order to prepare low-entropy states starting from product states with dopants pinned by anti-trapping potentials in the underlying Bose-Hubbard Hamiltonian. Because of the strong staggered magnetic field, we only consider fast parameter quenches and analyze dynamics starting from simple product states in the following.

\begin{figure}
\centering
\includegraphics{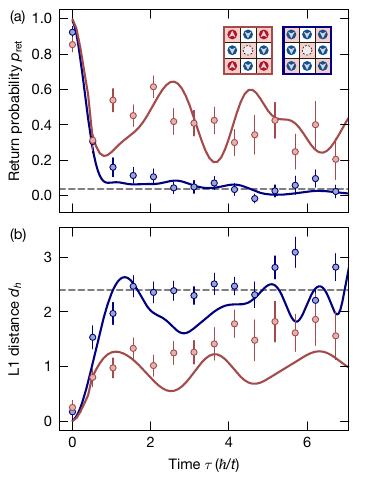} 
\caption{
\textbf{Experimental results for the spreading of a single hole in an AFM bosonic $t-J$ model with a strong staggered magnetic field}. We prepare a localized hole in the center of the system and inside different spin backgrounds: polarized (blue) and N\'eel (Eq.~\eqref{eqDefNeel}, red). In (a) we show the return probability and in (b) the L1 distance as a function of evolution time $\tau$, for parameters described in the text. Error bars denote the standard error of the mean. 
We compare to theory predictions (solid lines), where for the N\'eel state we perform time-dependent matrix product state simulations of the $t-J$ model on a $16 \times 8$ cylinder with $h_z/t=1.0$ and $t/J=5$. For the spin-polarized background state, we simulate the dynamics of a single particle in a $11 \times 11$ box with a staggered potential of $\pm h_z/4$, and analyze the hole density in a $5 \times 5$ box around the origin as in the experiment. The gray dashed lines in (a) and (b) indicate the values obtained by assuming a uniform density across the entire ROI.
}
\label{fig:exp_results}
\end{figure}

\section{Experimental analysis of single hole dynamics}\label{sec:exp_results}
Our experiments start with a spin-polarized Mott insulator of about 250 atoms with an average filling of $0.95$ in the cloud center.
By first addressing a central site and following an optical push-out, we prepare a single-hole initial state.
By subsequently applying staggered addressing, we can choose to change the polarized spin background to a N\'eel state, Eq.~\eqref{eqDefNeel}, with up to $\SI{98(3)}{\percent}$ fidelity per site, see Appendix~\ref{sec:exp-details}.
We then initiate dynamics by quickly lowering the lattice depth to a value corresponding to a hopping energy of $t = h \times \SI{11.8(5)}{Hz}$ and a Hubbard interaction of $U = h \times \SI{250(20)}{Hz}$, corresponding to a spin-exchange energy of $J = h \times \SI{2.2(2)}{Hz}$, while keeping a staggered field of $h_z / t = 2.0(2)$.
After the evolution time $\tau$, we suddenly ramp up the lattice depth for imaging and extract the atomic occupations.

In Fig.~\ref{fig:exp_results}, we show the spreading dynamics of a single hole in different spin backgrounds by analyzing the probability of the hole to return to its initial site, 
\begin{equation}
    p_\text{ret} (\tau) = \avg{\hat{n}_h (0, 0; \tau)},
\end{equation}
and the L1 (or Manhattan) distance, 
\begin{equation}
d_h (\tau) = \avg{\hat{n}_h (x, y; \tau) \cdot (|x| + |y|)}, 
\label{eqDefdh}
\end{equation}
calculated over the central $5 \times 5$ sites of our experimental system; here $\hat{n}_h (x, y) = (1-\hat{n}_{(x, y),\uparrow})(1-\hat{n}_{(x, y),\downarrow})$ denotes the hole density at site $(x, y)$.
We post-select snapshots to at most three holes within this box, as additional undesired holes (e.g. due to state preparation errors) may affect the signal.

In Fig.~\ref{fig:exp_results}(a) we find that the return probability quickly vanishes in the case of a spin-polarized background, saturating at a level comparable to the noise expected from undesired holes created during state preparation. This observation is in agreement with theoretical calculations assuming a coherent spreading of a single, initially localized particle (i.e. the hole) in a staggered potential $\pm h_z/4$. In contrast, after a common initial drop corresponding to nearest-neighbor hopping, a significant non-zero return probability $p_{\rm ret} \sim 0.4$ remains detectable at all times for the hole in a N\'eel background, indicating that the latter remains localized around the origin. Since the on-site potential disorder is expected to be weak and similar in both spin backgrounds, our measurements demonstrate that extended spin-correlations in the N\'eel state cause the observed localization of the doped hole. In particular, the comparison to the spin-polarized case -- in which the doped hole also experiences a staggered potential shifting neighboring sites out of resonance -- demonstrates that the staggered magnetic field $h_z$ itself is not sufficient to explain the observed localization of the hole in the N\'eel state; instead, the properties of the spin background play an important role.

The type of ``auto-localization'' of a doped hole in a 2D N\'eel state that we observe has been theoretically predicted already before the discovery of high-$T_c$ cuprate superconductors: Bulaevskii et al.~\cite{Bulaevskii1968} considered a $t-J_z$ model with pure Ising interactions stabilizing the product N\'eel ground state. They argued that the motion of a doped hole displaces the spins along its path, leaving behind a string of length $\ell$ of flipped spins and thus creating a memory of the hole's trajectory through the lattice. This moreover leads to a cost in potential energy 
\begin{equation}
    E_{\rm pot}(\ell)= \sigma~ \ell - \frac{J}{2} \delta_{\ell,0},
    \label{eqEpotString}
\end{equation}
where the linear string tension $\sigma = J$ provides a strong confining potential localizing the hole around the origin, even when $t \geq J$ represents the largest energy scale. 

Although the model realized in our experiment features approximately $SU(2)$ invariant spin-exchange couplings, the strong staggered magnetic field $|h_z| > J$ leads to similar physics as in the $t-J_z$ model. On the one hand, the field pins the N\'eel order, thus breaking the $SU(2)$ symmetry; on the other hand, the field adds to the linear string tension which now reads
\begin{equation}
    \sigma = \frac{|h_z|}{2} + J.
\end{equation}
For our experimental parameters, $\sigma = h \times \SI{14(1)}{Hz}$ is comparable to the tunneling energy $t$, leading to strong confinement of the hole. 

As a further consequence of the confining force, well-defined string-vibrational resonances of the doped hole have been predicted to appear at higher energies~\cite{Bulaevskii1968,Shraiman1988,Liu1991,Brunner2000,Mishchenko2001,Grusdt2018tJz,Bohrdt2020_ARPES}. Such excitations can be created by the initially localized hole, and are expected to give rise to oscillatory features in the return probability of the hole that become more pronounced for increasing values of the string tension $\sigma$~\cite{Bohrdt2020_dynamics,Hubig2020,Nielsen2022}. In our theoretical calculations, see Fig.~\ref{fig:exp_results}(a), such long-lived oscillations are clearly visible for a hole moving in a N\'eel background. Here we performed time-dependent matrix product state (t-MPS) simulations of the $t-J$ model on a $16 \times 8$ cylinder, as described in~\cite{Bohrdt2020_dynamics} but starting from a product N\'eel state with a single hole initialized in the center (see Sec.~\ref{sec:pair} for more details).

At short times we obtain excellent agreement between theory and experiment, demonstrating that the observed initial ballistic spreading of the hole in the N\'eel state is correctly captured by our model. The pronounced oscillations related to vibrational resonances of the string, on the other hand, are quickly washed out in the experiment. We attribute this to experimental imperfections, in particular due to undesired additional holes created during state preparation and residual potential disorder. Nevertheless, experimentally we observe a statistically significant 
non-monotonous behavior of the return probability in the N\'eel background, see Fig.~\ref{fig:exp_results}(a), with an indication of oscillations as expected from the calculation. This behavior is in contrast to the case of a spin-polarized background, where the measured return probability is consistent with a monotonous decay towards zero.

Next we discuss the dynamics of the L1 distance, shown in Fig.~\ref{fig:exp_results}(b) and describing the spreading dynamics of the initially localized hole. Consistent with the signatures observed in the return probability, we find a faster spreading of the hole in the spin-polarized background as compared to the case with a N\'eel initial state. The comparison to theory in the polarized case shows that $d_h$ grows to a distance $d_h|_{\rm max} \approx 2.5$ consistent with delocalization of the hole over the analyzed $5\times 5$ area. The agreement with theory is less clear than in the case of the return probability, which we attribute to the stronger effect of undesired holes on the L1 distance $d_h$: Randomly placed holes can significantly enhance the L1 distance, in particular at short times, due to the fact that these excess holes start away from the origin. The return probability on the other hand is only affected by excess holes if they propagate to the origin during the time evolution, which is typically a small effect. We partly account for the effect of excess holes on the L1 distance by subtracting the average hole density per snapshot and normalizing the resulting pseudo-hole density to one in the experimental data, see Appendix~\ref{sec:exp-details}.

In the case of a N\'eel background, our theoretical t-MPS calculations on extended cylinders predict pronounced oscillations also in the L1 distance, see Fig.~\ref{fig:exp_results}(b), again due to vibrational excitations of the string created by the hole motion. Note that the numerical simulations start from an ideal initial state with exactly one hole. Excess holes, as present in the experiment, will wash out these oscillations. Experimentally we cannot conclusively identify such oscillations within error bars, again likely due to the imperfections discussed above. Instead of oscillations, we find that the L1 distance keeps growing slowly, following a fast initial rise. Since the measured L1 values in the N\'eel background are well below those of the polarized system, and the latter are limited by the finite extent of the $5 \times 5$ analysis region, we can rule out finite-size effects to explain this slow growth. This result is consistent with our observation of ``auto-localization'' of the doped hole in the return probability, but suggests a weak remaining delocalization mechanism.


\begin{figure*}
\centering
\includegraphics{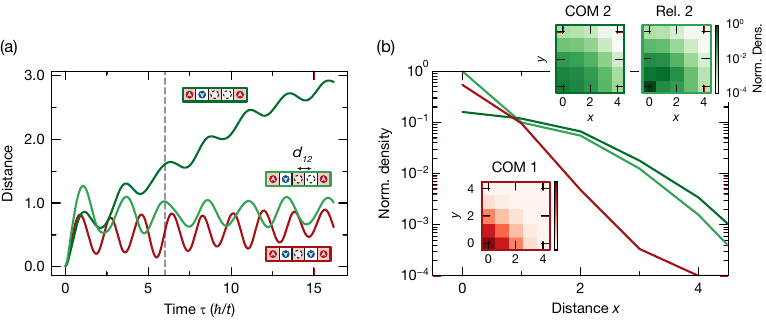} 
\caption{
\textbf{Theoretical comparison of one and two hole spreading}. Starting from the ground state of the undoped $t-J$ model on a $16\times 8$ cylinder and a large staggered field $h_z \gg t$, one (two) hole(s) are created by removing spins in the center. We analyze their dynamical delocalization, described by the bosonic, AFM $t-J-V$ model at $t/J=10$ and at a staggered field $h_z = 2t$. (a) Shows a fast ballistic spreading of the pair center of mass $d_{2h}$(dark green), whereas the single hole center of mass $d_{1h}$ (red) remains essentially localized around the origin, superimposed by a very slow ballistic spreading. Similar behavior, but without any detectable expansion at long times, is found for the L1 distance between the two holes $d_{12}$ (light green), indicating strong binding of the two holes.
(b) Shows the spatial profile of the symmetrized density distribution along the $x$-direction at $y=0$ for the three cases (same color-code as (a)) at time $\tau = 6 \hbar / t$ (vertical gray dashed line in (a)). The L1 distance between the first and the second hole follows the single-hole center of mass at short distances, and the two-hole center of mass for long distances. The insets show the full symmetrized 2d density profiles (edge color indicating the correspondence with lines).}
\label{fig:pair_spreading}
\end{figure*}

Theoretically, such a delocalization mechanism invalidating the idealized ``auto-localization'' picture for a doped hole in the Ising background was found by Trugman~\cite{Trugman1988}. Trugman identified loop configurations, where the hole encircles a plaquette one-and-a-half times, which allow the dopant to spread freely on one sublattice. However, given the small number of these loop-like trajectories compared to the exponential number of string states~\cite{Grusdt2018tJz}, the time-scale associated with Trugman-loop induced motion is long~\cite{Poilblanc1992,Chernyshev1999}: For parameters as in our experiments, we expect the L1 distance to grow by one lattice site on a time scale $\tau_{\rm Trug} \approx 80 \hbar / t$~\cite{Grusdt2018tJz}. This is still slower than the observed rate of growth of $\nicefrac{d}{d\tau} d_h \approx 1 / (15 \hbar / t)$, but close to the weak residual growth rate $\nicefrac{d}{d\tau} d_h \approx 1 / (60 \hbar / t)$ we find in our t-MPS simulations when $J \ll t$, see Appendix~\ref{ApdxSnglHleSpr}.  

In addition to Trugman loops, the hole can delocalize through spin-exchange processes $\propto J$~\cite{Kane1989,Sachdev1989,Grusdt2019}. Their effect on the hole motion is fully taken into account in our t-MPS simulations shown in Fig.~\ref{fig:exp_results}(b), at the experimentally relevant value of $t/J=5$. As a result, we find faster spreading of the hole at longer times, see Appendix~\ref{ApdxSnglHleSpr}. Up to the experimentally accessible times, our measurements in Fig.~\ref{fig:exp_results}(b) suggest a slightly faster delocalization of the hole, which cannot be explained by the effects of spin-exchange processes alone. We attribute this to residual undesired holes as well as thermal fluctuations of the spin background that were shown to enable a significant speed-up of the hole dynamics in a N\'eel background~\cite{Hahn2022}.

\section{Theoretical analysis of pair dynamics}\label{sec:pair}
Similarly to the formation and spreading of one hole, revealing the structure of a single dopant, the dynamical properties of a pair of two holes constitute an important problem, providing insights into the structure of bound states. Akin to the creation and subsequent release of a single hole as described above, the perhaps most straightforward approach is to create and release two holes on neighboring lattice sites in the ground state, or a low temperature state, of a spin background. 
For our experiment, the currently achievable homogeneity and average densities prohibited a detailed study of the two-hole dynamics and binding.
We therefore provide a theoretical analysis of the two-hole dynamics next, in the context of the bosonic, AFM $t-J-V$ model and in the presence of a strong staggered magnetic field -- as in the single-hole case analyzed so far. 

To this end, we performed large-scale numerical t-MPS simulations on a $16 \times 8$ cylinder using the TeNPy implementation~\cite{Hauschild2018,Hauschild2019} and the MPO based $W^{(II)}$ time evolution method~\cite{Zaletel2015}. The strong staggered fields suppress quantum fluctuations in the spin background, rendering the initial state close to a product state and helping to keep the overall entanglement low. This enables us to reach long evolution times despite the large width of $8$ sites of the cylinder we use in our simulations. We numerically calculate the time evolution of the system under Hamiltonian~\eqref{eqtJModel}, starting from the initially undoped ground state with a strong staggered field of $h_z/t=10$. In the latter, we create a pair of neighboring holes on the central two sites at time $\tau=0$. Simultaneously, we instantly ramp down the staggered field to a value $h_z/t=2$, which we keep fixed during the subsequent time evolution. This procedure mimics most closely a realistic experimental protocol. 

From the numerically computed MPS, we calculate the two-hole L1 (or Manhattan) distance, defined relative to the center of the system,
\begin{equation}
    d_{2h}(\tau) = \avg{\hat{n}_h(x,y;\tau)\cdot \left( \text{min}[ |x|,|x-1| ] +|y|
    \right)},
\end{equation}
which describes the spreading of the center-of-mass of the pair. Note that the distance is defined with respect to the closer position in $x$-direction, as the two holes are initially localized on sites $(0,0)$ and $(1,0)$. We then compare our results to similar calculations of $d_h(\tau)$ for only a single hole, see Eq.~\eqref{eqDefdh}.

Our results are shown in Fig.~\ref{fig:pair_spreading}. At very early times we find universal delocalization dynamics, which is identical for one and two holes. At around one tunneling time, this universality breaks down. In contrast to the single hole, which is ``auto-localized'' up to a slow expansion on a time-scale consistent with Trugman loop processes~\cite{Trugman1988,Grusdt2018tJz} and weak spin-exchange effects, we find that the two initially neighboring holes feature fast, ballistic spreading over time -- see the green curve in Fig.~\ref{fig:pair_spreading}(a).

To understand this striking difference between one and two holes propagating in a N\'eel background, we return to a discussion of the memory of the hole's trajectory imprinted into the surrounding AFM through the creation of strings of displaced spins. On one hand, this mechanism underlies the ``auto-localization'' of individual holes. On the other hand, in the case of \emph{two} holes in a N\'eel background, one hole can retrace the path of the first hole, and thereby place spins back onto the correct sublattice. This amounts to an erasure of the memory of the first hole's trajectory, and means that the center-of-mass of the pair can move freely through the AFM, i.e., without a memory of its trajectory -- similar to the free expansion of a single hole in a spin-polarized background. This phenomenon is at the heart of the fast observed spreading of a hole pair in Fig.~\ref{fig:pair_spreading}, and has been predicted previously to manifest in a broad, dispersive band in the two-hole spectrum~\cite{Shraiman1988,Grusdt2022,Bohrdt2023}.

In order to enable fast spreading of the pair, the two holes need to remain tightly bound to one another, allowing one hole to quickly remove the string created by the other; i.e., one hole needs to remain ``auto-localized'' around the second hole. This is achieved through the same mechanism that leads to ``auto-localization'' of individual holes around the origin: The potential energy of a string of displaced spins rises approximately linear with the length $\ell$ of the string, see Eq.~\eqref{eqEpotString}, providing a strong attractive confining force between the two holes~\cite{Bohrdt2022_bilayer,Grusdt2022}. In Fig.~\ref{fig:pair_spreading}(a) we find direct signatures of a strong confining force by analyzing the L1 distance $d_{12}(\tau)$ from one hole to the other (see also Fig.~\ref{fig:pair_spreading}(b)). Namely, $d_{12}(\tau)$ remains bounded at long times, closely following the behavior of the L1 distance $d_h(\tau)$ of a single hole, and much smaller than the extent of the center-of-mass coordinate $d_{2h}(\tau)$ of the pair. This proves the bound nature of the two-hole state at all numerically accessible times.

As in the single-hole case, Trugman loops and weak spin-exchange processes can lead to corrections beyond this picture: In the two-hole case they lead to string-breaking, causing the bound state of the two holes to decay into a pair of individual holes: Each of these remains approximately ``auto-localized'' and delocalizes through further Trugman loop or weak spin-exchange processes. Since the associated time scales are very long, we do not expect clear signatures of such processes in Fig.~\ref{fig:pair_spreading}. However, when the strength of the staggered field is reduced further, spin-exchange processes begin to dominate: They lead to a fast decay of the hole pair into two holes propagating through the AFM background individually -- each with a greater speed compared to the case with a strong staggered field, again due to spin-exchange processes providing an efficient delocalization mechanism. When spin-exchange processes dominate, the spreading dynamics of a single hole and a pair become comparably fast.
Hence, the strong staggered magnetic field plays an important role for observing two-hole spreading that is strongly distinct from individual one-hole dynamics in a N\'eel background.



\begin{figure}
\centering
\includegraphics{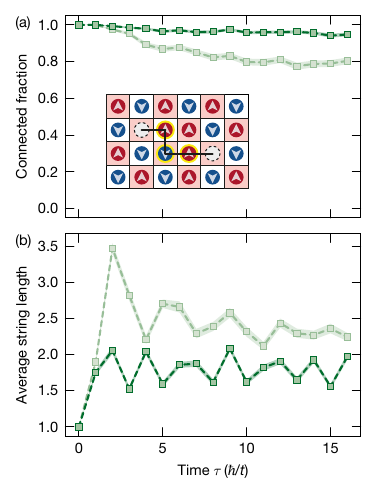}
\caption{
\textbf{String pattern analysis} between two dynamically spreading holes in the bosonic, AFM $t-J-V$ model, performed on snapshots sampled from numerically computed MPSs. We use $t/J=10$ and different strengths of the staggered magnetic field ($h_z/t = 2$ in dark green and $h_z/t = 1$ in light green). At each point in time $\tau$, exactly $80$ snapshots are evaluated, and shaded areas denote the standard error of the mean. In (a) we show the fraction of string patterns, as defined in the text, connecting the two holes (see inset for a specific example). In (b), the average length of the observed string patterns connecting two holes is shown.
}
\label{fig:strings}
\end{figure}

Next we analyze the two-hole bound state in more detail. Since the staggered magnetic field pins the N\'eel order of the system in $z$-direction, one can directly search for patterns of the strings connecting two holes in quantum projective measurements, or snapshots, of the quantum many-body state~\cite{Grusdt2018tJz,Chiu2019Science}. To this end, we define a string pattern in a given snapshot as a non-branching deviation from the underlying N\'eel state, where at least one site of the string pattern has to be adjacent to a hole, see inset of Fig.~\ref{fig:strings}(a). In the case of two holes, we can additionally consider the fraction of snapshots at a given time $\tau$ in which the two holes are connected by such a string pattern. 

In Fig.~\ref{fig:strings}(a), the probability $p_{hh}(\tau)$ to find a string pattern connecting the two holes is shown as a function of evolution time $\tau$. Initially, the holes are on nearest neighboring sites and thus always form a trivial string pattern. In the ensuing dynamics, the probability to find a connecting string pattern decays slowly, but stays large, e.g. around $90\%$ for $h_z/t=2.0$, on the time scales considered here. For weaker fields, e.g. $h_z/t=1.0$, also shown in Fig.~\ref{fig:strings}(a), the probability for finding a connecting string decreases. We attribute this to the increased influence of quantum fluctuations for weaker staggered magnetic fields, washing out the string pattern when the latter has formed after around one tunneling time. At longer times, when the distance between two holes has saturated, see Fig.~\ref{fig:pair_spreading}(a), we expect a very slow linear decrease of the connecting-string probability $p_{hh}(\tau)$ due to Trugman loop effects. This view is consistent with the observation that the slow observed time-scale in Fig.~\ref{fig:strings}(a) agrees with the slow time-scale associated with ballistic spreading of a single hole in Fig.~\ref{fig:pair_spreading}(a) that we previously associated with Trugman loop effects.

The average length of the string patterns connecting the two holes is shown in Fig.~\ref{fig:strings}(b), demonstrating that the holes form a rather tightly bound pair with an average size of two-to-three bonds, i.e., one-to-two sites between the two holes. As expected, the average string length is smaller for the stronger staggered magnetic field. It decays slowly in the case with a weaker staggered field $h_z/t=1$, which we again attribute to Trugman loop effects. Our analysis of string patterns between the two holes thus strongly supports the picture of a string binding mechanism between the two holes, leading to a tightly bound pair that can move freely through the AFM as a bound object.

\section{Summary and outlook}\label{sec:summary}
In this article, we have demonstrated an experimental scheme to realize the bosonic, AFM $t-J-V$ model in cold atom quantum simulators. To this end, we have used ultracold $^{87}\mathrm{Rb}$ atoms in a quantum gas microscope and leveraged negative absolute temperature states to realize antiferromagnetic couplings between bosons. This platform has furthermore enabled  the implementation of strong staggered magnetic fields through the use of single-site addressing techniques. On one hand, this allowed us to experimentally explore the textbook scenario of one-hole dynamics in a near-perfect N\'eel background, providing  evidence that Trugman loop and/or weak spin-exchange processes lead to slow delocalization of the ``auto-localized'' state of a single doped hole. Theoretically, we have pointed out the importance of a memory effect, where strong spin-charge correlations effectively create a memory of the hole's trajectory in the surrounding N\'eel state. On the other hand, the staggered magnetic field can be used in future studies for the adiabatic preparation of ultra-low entropy antiferromagnets and potentially the preparation of other low-temperatures states of the $t-J-V$ model, including at finite doping. This can be achieved by starting from a product state and adiabatically ramping the staggered field~\cite{Xie2022}. 

Second, we have numerically established a striking difference in the dynamics of a single hole versus a pair of neighboring holes in a N\'eel background subject to a strong staggered magnetic field. While the single hole remains approximately ``auto-localized'', we demonstrated fast spreading of two holes, which remain tightly bound to one another during this process. To observe such interaction-induced delocalization dynamics, the staggered magnetic field plays a crucial role: It stabilizes the surrounding N\'eel state and contributes to the linear string tension causing ``auto-localization'' and hole-pairing. Furthermore, it controls the quasiparticle residue $Z$ of the pair, i.e. the overlap of the paired ground state with the product state of two holes localized on neighboring sites that we start our dynamics from. A large staggered magnetic field ensures sizable values of $Z$, see Appendix~\ref{ApdxQpRes}, which in turn avoids decay from ro-vibrational string states~\cite{Shraiman1988,Grusdt2022} into unbound individual hole pairs through string-breaking processes. We find the latter effect to play a dominant role in the SU(2) invariant $t-J-V$ model \emph{without} a staggered magnetic field, where one- and two-hole spreading dynamics become challenging to discern.

While the current experiment we performed was limited by excess holes and disorder, there are no fundamental limitations to experimentally also probe the dynamics of a pair, or the finite doping regime.
Indeed, we expect that next-generation experiments drawing on advanced cooling techniques with global~\cite{Mazurenko2017} or local~\cite{Yang2020} entropy redistribution and better initial-state preparation will be able to study the predicted effects also for more than a single hole.
This potentially enables novel insights into strongly correlated systems: Our experimental scheme allows to disentangle the effects of fermionic statistics of the dopants, as we study bosonic charge carriers while keeping AFM interactions, and the interplay of spin and charge degrees of freedom. The latter can be studied systematically by tuning the strength of the staggered magnetic field.

\section{Acknowledgements}
We thank Tizian Blatz, Antoine Browaeys, Markus Greiner, Tim Harris, and Lukas Homeier for fruitful discussions. This work was funded by the Deutsche Forschungsgemeinschaft (DFG, German Research Foundation) under Germany's Excellence Strategy -- EXC-2111 -- 390814868. This project has received funding from the European Research Council (ERC) under the European Union's Horizon 2020 research and innovation program (Grant Agreement no 948141) — ERC Starting Grant SimUcQuam. This publication has also received funding under the Horizon Europe program HORIZON-CL4-2022-QUANTUM-02-SGA via the project 101113690 (PASQuanS2.1).
P.W. acknowledges funding through the Walter Benjamin program (DFG project 516136618)

\section*{References}

%

\newpage
\appendix

\section{Appendix}
\label{secAppendix}

\subsection{Experimental details}
\label{sec:exp-details}

In this section we provide more details on the experimental approach, characterize preparation fidelities, describe analysis procedures and discuss technical limitations.

\subsubsection{Experimental sequence}

\begin{figure*}
    \centering
    \includegraphics{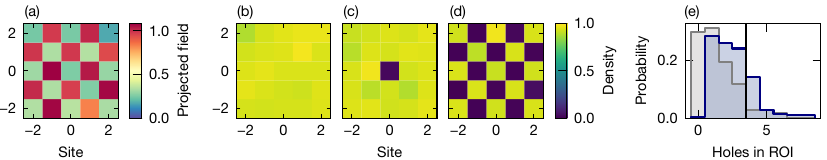} 
    \caption{
    \textbf{Experimental system preparation fidelity.}
    (a) Site-resolved spectroscopic measurement of the projected differential light shift for the staggered-field pattern, yielding a relative light shift on the addressed (non-addressed) sites of $h_\text{odd} = 1.00$ ($h_\text{even} = 0.29$) with standard deviation $\sigma_{h, \text{odd}} = 0.05$ ($\sigma_{h, \text{even}} = 0.06$).
    Average initial-state densities for (a) Mott-insulator preparation ($\overline{n} = 0.95(3)$), (b) single-hole preparation ($\overline{n}_\text{hole} = 0.02(1), \overline{n}_\text{atom} = 0.94(1)$) and (c) N\'eel-state preparation ($\overline{n}_\text{odd} = 0.01(1), \overline{n}_\text{even} = 0.94(1)$).
    (e) Probability distribution of the number of holes within the $5 \times 5$ ROI for the Mott insulator (gray) and single-hole state (blue). The black line represents the post-selection threshold, yielding a retaining fraction of $\SI{80}{\percent}$ for the hole preparation data set.
    }
    \label{fig:prep-fidelity}
\end{figure*}

Our quantum gas microscope employs $^{87}\mathrm{Rb}$ atoms trapped in a single layer two-dimensional layer of a retro-reflected vertical optical lattice and horizontally trapped in a bow-tie lattice with a square lattice spacing of $\SI{752}{nm}$.
We first prepared a spin-polarized Mott insulator of about $250$ atoms close to unity filling in the atomic limit, where tunneling is strongly suppressed (see Fig.~\ref{fig:prep-fidelity}(b)).
We then prepared the desired charge state by using a DMD to project light onto the central site, which generated a differential light shift between the spin states.
This allowed us to selectively flip the spin of the addressed atom with a microwave pulse, followed by a spin-selective optical push-out pulse to remove the addressed atom (see Fig.~\ref{fig:prep-fidelity}(c)).
The local differential-light-shifting approach furthermore allowed us to prepare 2D N\'eel spin states by projecting a checkerboard pattern onto the atoms, followed by a microwave pulse but no push.
To characterize the spin-state preparation, we again applied a spin-selective optical push-out pulse (see Fig.~\ref{fig:prep-fidelity}(d)).

For the projected light, we used $\sigma^+$-polarized laser light at a wavelength of $\SI{788.7}{nm}$, which corresponds to the calculated ``anti-magic'' wavelength for $^{87}\mathrm{Rb}$, where the two spin states are subjected to opposite light shifts.
This feature allowed us to also directly use the light pattern employed for the N\'eel state preparation as the staggered magnetic field during time evolution of our experiments.
By projecting a differential light shift of typically $\SI{20}{kHz}$, we could directly perform microwave spectroscopy to optimize the projected pattern (see Fig.~\ref{fig:prep-fidelity}(a)), resulting in residual field disorder of around $\SI{6}{\percent}$.
We additionally optimized the light polarization by maximizing the spectroscopically measured differential shift.
As the projected field required $\si{Hz}$-level (instead of $\si{kHz}$-level) light shifts, we stabilized the laser power using two photodiodes with different sensitivity, which we cross-calibrated using the high-power spectroscopic measurement.

As our measurements are highly magnetic-field sensitive, we furthermore probed for residual background fields using long-time Ramsey measurements in a deep lattice and minimized spin-dependent potential shifts due to magnetic-field gradients to less than $\SI{0.2}{Hz}$ per lattice site.

To initiate dynamics after the charge- and spin-state preparation, we first applied the staggered magnetic field and then ramped down the horizontal lattice from a depth of $28 E_r^{(752)}$ to $14.6(13) E_r^{(752)}$ in \SI{5}{ms}, where $E_r^{(a/\mathrm{nm})} = h^2 / 8 m a^2$ denotes the lattice recoil energy.
Following the time evolution, the horizontal lattice is ramped up to a depth of $80 E_r^{(752)}$ in \SI{1}{ms} to freeze the site populations and the atoms imaged without spin resolution.
The lattice depths were calibrated by lattice amplitude-modulation spectroscopy and the Bose--Hubbard parameters were obtained through bandstructure calculations.
The tunnelling energies along the two axes of the horizontal lattice during evolution featured an inhomogeneity of $\SI{5}{\percent}$ and the vertical lattice was kept at a depth of $12 E_r^{(532)}$ to minimize spatial inhomogeneities of the trapping potential, as discussed in Ref.~\cite{Wei2023}.

\subsubsection{Data analysis}

For the experimental data presented in the main text, we typically performed about $100$ experimental runs per data point prior to post-selection.
Although we were working deep in the $t-J$ model limit at $U/t = 21$, where quantum fluctuations of the atomic occupation in the Mott insulating phase are largely suppressed, finite temperature and spatial inhomogeneities led to the presence of holes in addition to the prepared hole.
As our signal is that of a single hole amidst these background holes, we decided to focus on a small $5 \times 5$ region of interest (ROI) in the center of the atomic cloud.
Additionally, we post-selected the measurements for a maximum of $3$ holes inside the ROI, chosen to correspond to a retention rate of about $\SI{80}{\percent}$ purely due to state preparation (see Fig.\ref{fig:prep-fidelity}(e)).
During time evolution, additional (parity-projected) holes formed in the atomic background due to the ensuing Hubbard dynamics.
This then led to a further reduction of the observed density, as shown in Fig.~\ref{fig:density-dyn}.

For the extraction of the hole propagation distance, we evaluated the hole density as follows:
To account for additional holes within the $L \times L$ ROI, for each snapshot we calculated the reference atomic density as $\overline{n}_\mathrm{ref} = \overline{n} L^2 / (L^2 - 1)$, where $\overline{n} = \sum_{x, y} n (x, y) / L^2$ is the average in-ROI density evaluated per snapshot.
We then define a pseudo-hole density as $n_h (x, y) \sim \overline{n}_\mathrm{ref} - n (x, y)$, where we choose the normalization such that $\sum_{x, y} n_h (x, y) = 1$.
We use this non-positive-definite definition of the hole density to obtain ROI-independent expectation values, unbiased by excess holes, for the experimentally extracted Manhattan distance.

\subsubsection{Density dynamics and potential inhomogeneities}

\begin{figure}
    \centering
    \includegraphics{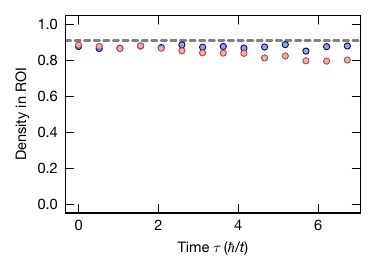} 
    \caption{
    \textbf{Experimental evolution of the density} (without post-selection of the hole number) within the $5 \times 5$ ROI for the N\'eel state (red) and the spin-polarized state (blue).
    The staggered magnetic field is oriented anti-parallel (high-energy state) to the N\'eel-state spins.
    The gray dashed line indicates the density expected from the state-preparation fidelity.
    }
    \label{fig:density-dyn}
\end{figure}

In our analysis of the hole propagation and the studied $t-J-V$ model, we assume the background atoms to remain in a stable unity-filled Mott insulator state.
To verify that the background atoms do not dominate our experimental signatures, we analyze the parity-projected average density in the ROI prior to any post-selection over time in Fig.~\ref{fig:density-dyn}.
In the polarized case, we observe that the density remains largely constant, as expected for the high-$U/t$ regime, which we are working in.
In contrast, the density in the high-energy N\'eel configuration decreases slightly with time, suggesting the presence of doublon--hole fluctuations facilitated by the staggered field (as the effective local nearest-neighbor potential difference is doubled with respect to the spin-polarized case).
This would not have a strong influence on the single-hole dynamics at the level of precision of the experiment, especially when post-selecting the data for a limited number of excess holes.
However, this effect leads to a significant reduction in the signal-to-noise ratio attainable and presently precludes us from direct measurements of two-hole dynamics.

Additionally, our system is subjected to a spin-independent spatial potential inhomogeneity, primarily induced by laser speckles of our vertical optical lattice.
This leads to potential variations on the scale of \SI{50}{Hz} over larger areas (typically on the order of $10$ sites), which can result in similar site-to-site potential variations as the energy scale of the staggered field, particularly away from the cloud center.
This can in principle be corrected by additional spin-independent light potentials, and constitutes a future improvement for the observation of multi-hole dynamics.

\subsubsection{Field strength}

\begin{figure}
    \centering
    \includegraphics{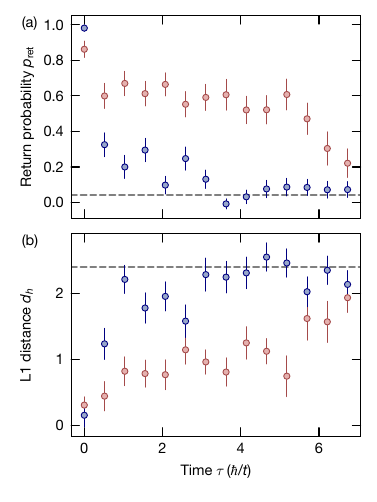} 
    \caption{
    \textbf{Experimental single-hole dynamics at higher field strength} of $h_z / t = 3.0(2)$.
        Similar to Fig.~\ref{fig:exp_results}, we show the hole's L1 distance (a) and return probability (b) over time for a N\'eel (red) and a polarized (blue) spin background. We again find localization only for the N\'eel state case and obtain the expected stronger localization.
    }
    \label{fig:loc-field-strength}
\end{figure}

To corroborate the single-hole dynamics presented in the main text and measured at a staggered-field strength of $h_z / t = 2.0(2)$, we repeated the measurement at a stronger field of $h_z / t = 3.0(2)$, see Fig.~\ref{fig:loc-field-strength}.
Here, we qualitatively find the same behavior, namely localization of the hole in a N\'eel state spin background, whereas the hole may delocalize in the spin-polarized case.
Quantitatively, we observe that the hole in the N\'eel background remains more strongly localized, as evidenced by the smaller L1 distance and larger return probability compared to the $h_z / t = 2$ case, which is in agreement with the expected increase in string tension with field strength.

\subsection{Single hole spreading}
\label{ApdxSnglHleSpr}

\begin{figure}
\centering
\includegraphics{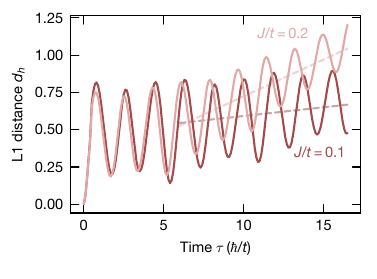}
\caption{
\textbf{Effect of spin-exchange $J$ and Trugman loops} on the dynamics of a single hole in the $t-J$ model with a strong staggered field $h_z/t = 2.0$ and spin exchange values of $J / t = 0.1$ (dark red) and $0.2$ (light red). A linear fit to evolution times $\tau > 6 \hbar / t$ (dashed lines) yields a velocity of $v = 0.012 t / \hbar$ and $0.049 t / \hbar$, respectively.
}
\label{fig:Trugman_spreading}
\end{figure}

The possibility to tune the spin-exchange $J$ at a fixed, large value of the staggered magnetic field $h_z/t = 2.0$ allows us to disentangle the effects of Trugman loops and spinon dynamics, see Fig.~\ref{fig:Trugman_spreading}. A linear fit to the data from time $\tau = 6 \hbar / t$ onwards yields a velocity of $v = 0.012 t / \hbar$ for $J/t = 0.1$, consistent with the spreading through Trugman loops discussed in the main text. For $J/t = 0.2$, the spinon dynamics itself starts to play a role and a linear fit to the numerical data yields $v = 0.049 t / \hbar$.


\subsection{Pair quasiparticle residue}
\label{ApdxQpRes}

\begin{figure}
\centering
\includegraphics{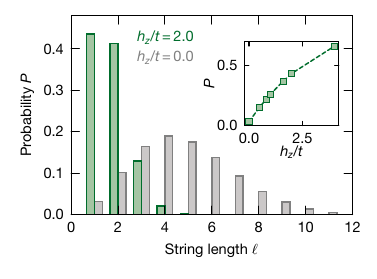}
\caption{
\textbf{String length distribution} of two holes in the ground state of the AFM bosonic $t-J$ model at $t/J=3$ with a staggered magnetic field from the geometric string theory~\cite{Grusdt2018tJz,Grusdt2019,Grusdt2022}. The probability for a string of length $1$, shown in the inset as a function of the staggered field strengths, corresponds to the overlap of the initial state of two holes next to each other with the ground state of the pair.
}
\label{fig:string_distribution}
\end{figure}

In the absence of the staggered magnetic field, the initial state with two neighboring holes has a rather small overlap with the ground state of the pair and thus constitutes a high energy state that can decay into two independently propagating magnetic polarons. In the ensuing dynamics, it is thus almost impossible to distinguish properties of pairs from two individual magnetic polarons. This can for example be seen from the spectral weight in the low energy branch of the two hole spectral function, which considers the creation of two holes on nearest neighboring sites as excitation~\cite{Bohrdt2023}. A complementary perspective is provided by geometric string theory~\cite{Grusdt2019}, which has proven quite successful in describing the properties of a single hole as well as the pair spectral function~\cite{Bohrdt2023,Grusdt2022}: in the SU(2) invariant $t-J$ model with $t \gg J$, the string length distribution for the string connecting two holes is centered around larger values of lengths, e.g., with a peak at $\ell \approx 3$ for $t/J=3$, see also Fig.~\ref{fig:string_distribution}. The initial state of two holes on nearest neighboring sites corresponds to a string of length $1$, which has a comparably low probability of $\approx 3\%$.

A much larger overlap of the initial state with two holes on nearest neighboring sites with the ground state of the pair can be obtained through a strong staggered magnetic field, see Fig.~\ref{fig:string_distribution}. The ensuing real-time and space dynamics of two dopants can thus directly provide insights into the stability, as well as the effective mass of the pair. 


The two holes, on the other hand, have a large overlap with the ground state of the pair, see also Fig.~\ref{fig:string_distribution}, and thus move together through the system. This can be seen by considering the distance between the two holes as a function of time, see Fig.~\ref{fig:pair_spreading}, which, after a fast initial rise, oscillates around a distance of two lattice sites. Thus for short times, the pair slightly extends from the nearest-neighbor bond on which it has been initialized, and subsequently spreads through the system. 
Note that the distance between the two holes is comparable to and indeed shows similar oscillations as the distance of a single hole to the origin, indicating that the underlying mechanism for binding the two chargons together is similar to the mechanism binding the spinon to the chargon in the case of the magnetic polaron. 
The pair of two holes can further spread through the system because, as opposed to the magnetic polaron, the compound object consists of two chargons and no spinon. Put differently, when the first hole moves, the second hole can retrace the string of displaced spins left behind and thereby restore the sublattice magnetization favored by the staggered field.

\end{document}